\documentstyle[aps,epsf,prl,multicol]{revtex}
\begin{document}
\draft
\title{ Chaos and Thermalization in a Dynamical Model of
Two Interacting Particles
}
\author{F.Borgonovi$^{[a,b,c]}$, I.Guarneri$^{[d,b,c]}$,
F.M.Izrailev$^{[d,e,f]}$ and G.Casati$^{[d,c,g]}$ }
\address{
$^{[a]}$Dipartimento di Matematica, Universit\`a Cattolica,
via Trieste 17, 25121 Brescia, Italy  \\
$^{[b]}$Istituto Nazionale di Fisica Nucleare, Sezione di Pavia,
via Bassi 6, 27100 Pavia, Italy\\
$^{[c]}$Istituto Nazionale di Fisica della Materia, Unit\`a di Milano,
via Celoria 16, 20130 Milano, Italy\\
$^{[d]}$  Universit\'a di Milano, sede di Como,
via Lucini 3, 22100, Como, Italy\\
$^{[e]}$Budker Institute of Nuclear
Physics, 630090, Novosibirsk, Russia\\
$^{[f]}$Instituto de Fisica, Universidad Autonoma de Puebla, Apdo.
Postal J-48, Puebla, 72570 Mexico\\
$^{[g]}$Istituto Nazionale di Fisica Nucleare, Sezione di Milano,
via Celoria, Milano, Italy\\
}
\date{\today}
\maketitle 
 
\begin{abstract}
A quantum dynamical model of two interacting spins, 
with  chaotic and regular components, is investigated
using a finite two--particles symmetrized basis.
Chaotic eigenstates give rise to an equilibrium 
occupation number distribution 
in close agreement with the Bose--Einstein distribution
despite the small number of particles ($n=2$).
However, the corresponding temperature differs from that 
derived from the standard Canonical Ensemble.
On the other side, an acceptable agreement with the latter is restored by  
artificially randomizing the model.
Different definitions of temperature 
are then discussed and compared . 
\end{abstract}
 
\begin{multicols}{2}

The concept of quantum ergodicity, and its connections with the 
foundations of statistical mechanics,  have a long history which, in the 
last years, has received new impulse from the investigation of quantum
systems which are chaotic in the classical limit.  
The important issue here is twofold. First, one would like to know 
if such quantum systems  display some sort 
of chaotic behaviour; second, if this behaviour will provide foundations 
for quantum statistical mechanics. In other words, the problem is  whether 
a $N$- body isolated
system, which, on the classical side, displays sufficiently strong 
chaotic properties, will, on the quantum side, give rise to 
conventional statistical properties, such as, e.g., the 
 Bose-Einstein (BE) or Fermi-Dirac (FD) 
distributions\cite{cagu}.
The first problem has been widely investigated for single-particle 
systems; not so for many-body systems, essentially because such systems 
are not easily accessible to numerical investigations. 
 Instead,  
little is known about the second problem.

Whereas early studies on the foundations of quantum 
statistical mechanics did not attach special importance to  the structure of 
eigenstates, quantum ergodicity is nowadays 
associated with eigenstates being
homogeneously spread, in an appropriate  statistical sense, over the whole
energetically allowed range. 
Much of our current understanding of the structure of eigenstates for 
classically chaotic systems is based on the study of single-particle systems,
and on the analysis of eigenfunctions of Random Matrices, which have been 
often used as models for quantum chaos. In particular, suitably constructed 
Wigner Banded Random Matrices (WBRM) \cite{W55} are conjectured to preserve 
many of the spectral properties of "chaotic" Hamiltonian matrices \cite{FLP89}.
For WBRM, several results are known 
\cite{FLW91,CCGI93,FCIC96,CCGI96}, concerning spectrum statistics, 
structure of eigenstates and of the Local Spectral  Density of States 
(also called Strength Function),  
and conditions for quantum ergodicity have been given \cite{CCGI96}. These 
results provide paradigms of quantum ergodic behaviour, which should be 
tested on realistic Hamiltonians. 
 
If quantum ergodicity can be established along such lines for quantum 
many-body systems, then the problem arises, whether it gives rise to some 
sort of statistical equilibrium.  One would like to know whether the 
quantum averages of the occupation 
numbers of single-particle energy levels, taken over 
many-body ergodic eigenstates (or, 
more properly, over mixtures of eigenstates with energies lying in a narrow 
selected range), yield some statistically stable distribution law; if this is 
the case, how large must the number of particles be, in order that this 
distribution reproduces the conventional Fermi-Dirac or Bose-Einstein  
statistics?

This theoretical approach has been advocated in 
Refs.\cite{FIC96,FI97,FI97a};  
in this Letter we numerically 
investigate the just sketched theoretical issues  on an isolated system of  
 two interacting particles, which, under appropriate conditions, is 
classically chaotic in some energy region. 
This model has been proposed and investigated in  \cite{FePe} within the 
framework of quantum chaos; here we analyze it from the standpoint of 
statistical mechanics, and show that thermalization occurs 
in the classically chaotic energy region, 
in a sense that will be discussed below. 
 
We first review some fundamental facts about its classical and quantum 
behavior.
The model describes  two coupled rotators, with angular momentum $\vec{ L}$
, $\vec{ M}$ and Hamiltonian :
\begin{equation}
H  =  H_0 + V = (L_z + M_z ) +  L_x M_x 
\label{ham}
\end{equation}

It may be used to describe
 the interaction of quasi-spins in nuclear physics.
Constants of motion   are $H=E$, $L^2$ and $M^2$. 
It is worth to mention that in this form the dynamical 
variables $\vec{L}, \vec{M}$ are not canonical.

The analysys of the  surfaces of section reveals a large number of 
regular trajectories covering invariant tori when $L^2, M^2$ are both
very small or very large\cite{FePe}. 
To simplify the problem we set $L=M$.
In such a case the most interesting situation occurs when $1<L<10$
where, depending on the energy value $E$, regular and chaotic  
regions coexist. Typically when $\vert E \vert $ is close to 
the maximum allowed energy $E_{max}= L^2 +1$\cite{FePe} 
trajectories are regular while for $E\simeq 0$ the islands of stability 
become very small and chaotic motion dominates.
 
Quantization follows  standard rules, and angular momenta
are quantized according to the relations 
$L^2 = M^2 = \hbar^2 l(l+1)$ 
  where  $l$ is an integer number.
Therefore, for given $l$   the Hamiltonian is a finite matrix, 
and the semiclassical limit is recovered in the limit  $l \to \infty$
and $\hbar \to 0$  keeping $L^2$   constant.

The matrix elements 
 in the basis  $\vert l_z, m_z \rangle$
have the form,

\begin{eqnarray}
\langle l_z', m_z' \vert
H_0  \vert l_z, m_z \rangle  =
\delta_{m_z, m_z'}  
\delta_{l_z, l_z'} \hbar (l_z +m_z) &\nonumber  \\  
\langle l_z', m_z' \vert V   \vert l_z, m_z \rangle = 
{ {\hbar^2}\over {4}} 
 \delta_{m_z, m_z' \pm 1}   \delta_{l_z, l_z' \pm 1}\times & 
\label{hint}  \\ 
\times [(l+l_z) (l-l_z+1) (l+m_z) (l-m_z +1)]^{1/2} & \nonumber\\  
\nonumber
\end{eqnarray}
with $l_z, m_z$ integers,
 $-l \le l_z,m_z \le l$ .

The $z$--component of the 
total angular momentum $J_z = L_z + M_z$ (which is the 
unperturbed Hamiltonian $H_0$)
obeys the selection rules
$\Delta J_z = 0, \pm 2\hbar$, so the subspace spanned by 
the states with  odd $J_z$  
can be separated from  that with  $J_z$ even (there are no matrix
elements for the transition between them).
In what follows, we fix    $J_z =H_0 $ even (multiple of $\hbar$).

A key point in our approach is to represent the 
Hamiltonian in the symmetrized two--particle basis of non--interacting
particles.
This corresponds to the well-known ``shell model'' representation
used in  atomic and nuclear physics. 
Here, 
 we  restrict 
our considerations only to symmetric states with respect to the exchange 
of the two particles. 
In the symmetrized basis  
each set of states  with   fixed even $H_0$   
has a degeneracy
$l+1 - \vert H_0 \vert / 2 \hbar$, and 
the dimension of the Hamiltonian 
matrix is
$N =(l+1)^2$.
 Finally,  we reorder the matrix according to increasing unperturbed
energy, and thus obtain 
 a  band matrix,   
with high sparsity within the band (each line has at most 5 elements).

Direct diagonalization of such a matrix gives 
the  eigenfunctions $\psi_n (E_m)  $ 
of the total Hamiltonian $H$ represented in the ordered
symmetrized  two-particle basis $\vert {\bf n} \rangle$. 
Here $\psi_n (E_m)$ is the $n$-th component of the eigenfunction 
having $E_m$ as eigenvalue.
A  detailed analysis  of the structure of eigenstates, 
(which  will be reported in detail 
elsewhere\cite{borgo}) reveals that eigenstates which belong to 
the classical chaotic region are ergodic, in the sense that they fill a range 
of unperturbed energies, in a way which corresponds to the classical 
microcanonical distribution.  
Here we concentrate 
on the distribution of  the
occupation numbers of single-particle states. 

\begin{figure}
\vspace{1cm}
\hspace {-0.2cm}
\epsfxsize 8cm
\epsfbox{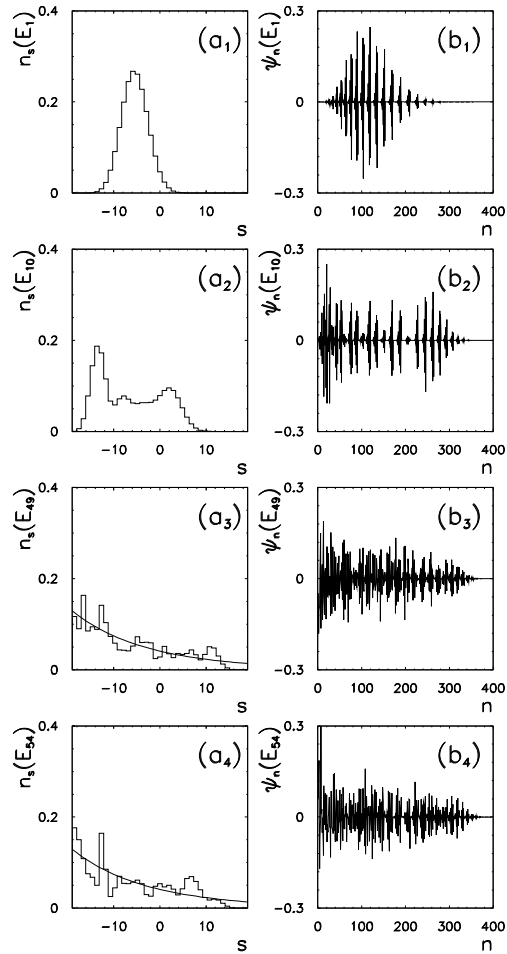}
\vspace{1cm}
\narrowtext
\caption{
Eigenfunctions (right column) and the corresponding occupation
number distributions (left column) for the case $L=3.5$, $l=19$.
($a_1$) and ($b_1$) are for the ground state,
($a_2$) and ($b_2$) are for the $10^{th}$ state (classically quasi-integrable),
($a_3$) and ($b_3$) are for the $49^{th}$ state (with a chaotic phase space
and $e_m = E_m/E_{max} =-0.464$) and the 
full line is the best fit to the BE distribution 
with $\beta_{fit}=0.305$;
($a_4$) and ($b_4$) are for the $54^{th}$ state (classically
chaotic) with 
$e_m =-0.449$,
$\beta_{fit}=0.310$
}
\label{bef}
\end{figure}

The distribution $n_s$ of occupation  numbers of single particle levels  $s$ 
can be directly obtained
from eigenfunctions. Given an eigenfunction
$\psi_n (E_m)$ one can write: 

\begin{equation}
 n_s ( E_m)  =  \sum_{n} \vert \psi_n (E_m) \vert^2
\langle {\bf n} \vert \hat{n}_s \vert {\bf n} \rangle  
\label{occup}
\end{equation} 

where $\hat{n}_s$ is the occupation number operator. 
The term  $
\langle {\bf n} \vert \hat{n}_s \vert {\bf n} \rangle  
$ equals $0,1,2$ depending on how many particles are located on the 
specific single-particle level $s$.
In Fig.\ref{bef} some examples for the occupation numbers
distribution (histograms on the left column) are given, together with 
the corresponding eigenstates (right column).

One can see a strong difference
between the distributions obtained from
"regular" eigenstates, belonging to the stable region ($b_1$,$b_2$),
and those obtained from ergodic ones, belonging to
the chaotic region ($b_3,b_4$).
In the stable region close eigenstates in energy $E_m$ yield
completely different distributions $n_s$
(see  Fig.\ref{bef} ($a_1,a_2$)), which means that there is no equilibrium
in the statistical sense\cite{FI97a}.

In contrast, in the classically chaotic region
 the form of the distribution $n_s$ is
statistically stable with respect to the choice of
a specific chaotic eigenstate (compare  Fig.\ref{bef} $a_3, a_4$).

\begin{figure}
\epsfxsize 6cm
\vspace {-0.5cm}
\hspace{1cm}
\epsfbox{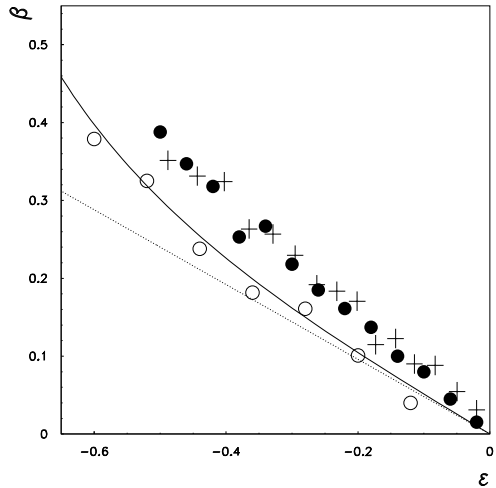}
\vspace {-0.5cm}
\narrowtext
\caption{
Dependence of the different definitions of
inverse temperature $\beta$  on 
the rescaled energy $\varepsilon = E/E_{max}$;
full curve: $\beta_c$, dotted curve : $\beta_T$,
full circles : $\beta_{fit}$, crosses : $\beta_{BE}$.
Open circles are $\beta_{fit}$ for the random model.
 }
\label{betae}
\end{figure}

In the standard thermodynamical treatment, temperature can 
be defined in a number of different ways, which are known 
to be equivalent in the thermodynamical limit.
It is interesting to compute temperature for our system,
following these different definitions, although our system
is far from  this limit. 
First of all we use the canonical expression: 

\begin{equation}
\langle E \rangle_{\beta_c} = { 
{\sum_m E_m e^{-\beta_c E_m}}
\over 
{\sum_m e^{-\beta_c E_m} } 
}
\label{can}
\end{equation}

where $E_m$ are the exact eigenenergies of the interacting system.
The above relation between energy and temperature
allows for standard thermodynamical description of our system.
The solution $\beta_{c} (E)$ of equation (\ref{can})  
 is shown  in Fig.\ref{betae} as a full curve.

For a system of $N$ non--interacting particles
with total energy $E$, the canonical
distribution is well--known to result,
when $N$ is large, in the BE distribution for the   
occupation numbers: 
$n_s^{BE} =   
[  e^{\beta(\hbar(s+l)-\mu)} - 1 ]^{-1}
\ \ (s=-l,\dots,l)$
where $\mu$ is the chemical potential and $\beta$ 
is the inverse temperature.
This expression is derived for an ideal gas
(many non-interacting particles) in   contact with a thermostat; 
in contrast, 
our system is isolated, with  two interacting particles only. Nevertheless, 
recent analytical and numerical
studies  for  random two-body
interaction \cite{FIC96,FI97,FI97a}, 
suggest that conventional quantum statistics can appear even in isolated 
systems
with relatively few particles, provided a proper renormalization of energy 
is taken.

As a   simple comparison, we may consider the BE 
distribution  
as one-parameter fitting expression, taking into account
the constraint set by the finite number of particles ($\sum_s n_s^{BE} = 2$).
This allows to find 
the corresponding inverse 
temperature $\beta_{fit}$, and from that the BE distribution 
presented in the left 
column of Fig.\ref{bef} by full curves. 

A different way of comparing numerical data with the  BE distribution
is to solve the following equations in the unknowns $\beta,\mu$:

\begin{equation}
\sum_{s=-l}^l n_s^{BE} = 2  \ ,  \ 
\sum_{s=-l}^l \hbar s \ n_s^{BE} =   {\cal E }
\label{norm}
\end{equation}

with $ {\cal E} = \sum_s \hbar s n_s$, 
computed from 
numerical values of $n_s$ (note that ${\cal E}$ is different from the 
exact eigenenergy because our model is strongly non--perturbative\cite{borgo}).
Doing so, and averaging over a number
of chaotic eigenstates with close values inside
 small energy windows, we have 
found inverse temperatures $\beta_{BE}$ quite close to the previously 
obtained  $\beta_{fit}$ 
(compare crosses with full circles in Fig.\ref{betae}).
The agreement  between the numerical values 
of $\beta_{fit}$ and $\beta_{BE}$  
supports, on one hand,  the significance of the fitting procedure with 
a BE distribution and, on the other hand, 
the validity of the BE distribution for isolated systems with few interacting
particles via  a proper renormalization of the energy $ {\cal E} $, see
\cite{FI97,FI97a}.

We have then compared $\beta_{fit}(E)$ with $\beta_c(E)$, where $E$ 
is the exact 
eigenenergy, and have found them to be significantly different.
This is of course hardly surprising: even in the presence of 
ergodicity, with so few particles, one cannot  expect coincidence of 
microcanonical and canonical averages, which still strongly resent of the 
particular choice of the interaction. In classical terms, the distribution 
of single-particle energy still depends on  the particular 
shape of the energy surface, which is in turn determined by the particular 
interaction chosen. 

A remarkably different result was obtained on 
"randomizing" our model, by  replacing non zero  
off-diagonal matrix elements in the Hamiltonian by random variables with the 
same mean and variance as in the exact dynamical model (\ref{ham}). 
Whereas $\beta_c$ is left practically unchanged by this replacement, 
$\beta_{fit} \simeq \beta_{BE} $ considerably changes, and it comes quite close  
to $\beta_c$. In our understanding, the reason of this striking result is that 
the random model (which has no smooth classical limit) uses 
a much more "generic" form of the perturbation than the dynamical one; 
in addition, the 
dependence on the specific interaction is further weakened by the 
(matrix) ensemble averaging. As a result, our data indicate that 
the conventional canonical distribution 
of occupation numbers may appear even in an 
isolated system with a quite 
small number of particles with random interactions, where the dynamical 
correlations, which prevent a similar results in the classical model, are 
negligible.

Other different definitions of temperature have been widely
discussed in application to complex nuclei \cite{Zele} in the
context of the onset of thermalization. It was in particular shown
that these definitions of temperature give the same result 
in realistic shell models of nuclei.
This fact was connected with the onset of ``true'' 
thermalization. However, as was
shown in \cite{FI97a}, when the number of interacting particles is small, 
one can
get different values of temperature depending on the particular definition,
even in the equilibrium region, where a statistical
description is legitimate. 
In this connection we have also compared the temperatures found from the 
BE distribution with the standard thermodynamical inverse
temperature defined by, 
$ \beta_T =  d \ ln \rho/ d E $
where $\rho$ is the density of states of the total Hamiltonian. 
Both for the dynamical and the random model, a Gaussian fit provides 
an excellent approximation to actual data for the density of states
\cite{Bohigas}. We 
have used this very fit in computing $\beta_T$, 
obtaining practically the 
same 
result in both cases. The function $\beta_T (E)$ is different from 
$\beta_c(E)$, but it comes closer and closer to it near the center of the 
spectrum (where both temperatures are infinite). For the dynamical model, 
$\beta_T(E)$ 
turns out to be completely different from both $\beta_c$ and $ \beta_{fit}
\simeq \beta_{BE}$ (see  dotted line in Fig.\ref{betae}).
This means that for small number of particles the above
definition  of $\beta_T$   is irrelevant to the distribution 
of occupation numbers\cite{FI97a}. For the random model, $\beta_T$ is not far 
from $\beta_{fit}(E)$ at high temperatures (where $\beta_T$ and $\beta_c$ 
tend to coincide), but deviates from it at smaller temperatures.
We do not push our comparison of $\beta_{fit}(E)$ and $\beta_c$ to smaller 
temperatures than shown in Fig.\ref{betae}, because near the edges of the 
spectrum, where the density of states is small,
 the eigenstates are not fully chaotic any more (note that in the dynamical 
model the motion becomes increasingly regular there).

In conclusion, we have studied a 
dynamical model of two--interacting Boson particles 
in a  finite
dimensional Hilbert space.
We have shown that, in the shell model representation,
the structure of exact eigenstates can be directly related to the onset of
equilibrium for the occupation numbers $n_s$ of single--particle states.
Specifically, for the eigenstates corresponding to classical
chaotic motion, an equilibrium distribution for $n_s$ occurs
which allows for a statistical description of the model.
In contrast, ``regular'' eigenstates  results in extremely 
non--generic fluctuations of $n_s$ for small changes of the energy, 
thus invalidating any  statistical approach.
For chaotic eigenstates, 
the distribution of occupation numbers can be approximately described
by the Bose-Einstein distribution, although the system is
isolated and consists of two particles only.
In this case a strong enough interaction plays the role of
a heat bath, thus  leading to thermalization. In spite of this, 
the minimal number of particles prevents 
the canonical distribution from describing our dynamical system  even if 
a surprising agreement with the canonical distribution is recovered 
in the corresponding random model.

  The authors are grateful to V.Flambaum for fruitful discussion.
  (FMI) thanks with pleasure the colleagues of the University of Milan
  at Como for the hospitality during his visit when this work was done;
  he acknowledges the support from the Grant of the Cariplo Foundation
  for Research and partial support from the INTAS Grant No. 94-2058.

\end{multicols}
\end{document}